\documentclass[11pt,a4paper]{article}
\usepackage{moriond}
\usepackage{graphicx}
\usepackage{natbib} 
\usepackage{amstext}
\usepackage{amssymb}
\usepackage[latin1]{inputenc}
\usepackage{sidecap}
\usepackage{aas_macros}

\def\be{\begin{equation}}
\def\ee{\end{equation}}
\def\bea{\begin{eqnarray}}
\def\eea{\end{eqnarray}}

\newcommand{\ion}[2]{#1~\textsc{#2}}
\newcommand{\sect}[1]{Sect.~\ref{#1}}
\newcommand{\arcsec}{\ensuremath{''}}
\newcommand{\HeI}{\ion{He}{i}}
\newcommand{\HeII}{\ion{He}{ii}}
\newcommand{\Brg}{Br$\gamma$}
\newcommand{\Paa}{Pa$\alpha$}
\newcommand{\lwl}[2]{#1~$\lambda$~\microns{#2}}
\newcommand{\microns}[1]{\ensuremath{#1\text{~}\mu\text{m}}}
\newcommand{\kms}[1]{\ensuremath{#1\text{~km~s}^{-1}}}
\newcommand{\SgrA}{Sgr~A}
\newcommand{\SgrAstar}{\SgrA$^\star$}
\newcommand{\SgrAWest}{\SgrA\ West}
\newcommand{\solarmasses}[1]{\ensuremath{#1\text{~M}_\odot}}
\newcommand{\Av}{\ensuremath{A_V}}

\begin{document}
\vspace*{4cm}
\title{CENSUS OF THE GALACTIC CENTRE EARLY-TYPE STARS USING SPECTRO-IMAGERY}

\author{T.~PAUMARD  (1),   R.~GENZEL  (1),  J.P.~MAILLARD   (2),  T.~OTT  (1),
  M.~MORRIS (3), F.~EISENHAUER (1), R.~ABUTER (1)}

\address{(1) Max-Planck-Institut f\"ur extraterrestrische Physik\\
Giessenbachstrasse, Postfach 1312, D-85741 Garching, Germany}
\address{(2) Institut d'astrophysique de Paris\\
98bis bvd. Arago, F-75014 Paris, France}
\address{(3) University of California, Los Angeles\\
  Div. of Astronomy, Dept of Physics and Astronomy\\
  Los~Angeles, CA 90095-1562, USA}

\maketitle\abstracts{The  few  central parsecs  of  the  Galaxy  are known  to
  contain a surprising  population of early-type stars, including  at least 30
  Wolf-Rayet  stars and luminous  blue variables  (LBV), identified  thanks to
  their strong emission lines.  Despite  the presence of emission from ionised
  interstellar   material  in  the   same  lines,   the  latest   advances  in
  spectro-imaging have made it possible to  use the absorption lines of the OB
  stars to characterise them as  well. This stellar population is particularly
  intriguing in the deep potential well of the 4 million solar mass black hole
  \SgrAstar. We  will review  the properties of  these early-type  stars known
  from spectro-imagery, and discuss possible formation scenarios.  }

\section{Introduction}
The  Galactic  Centre (GC)  is  the  closest of  all  galactic  nuclei in  the
universe. It is  also a galactic nucleus which shows  some traces of activity:
it  features  one  of  the  best supermassive  black  hole  (SMBH)  candidates
(\SgrAstar), the  densest star  cluster in the  Galaxy, an  \ion{H}{ii} region
(\SgrAWest\  or   the  Minispiral),  and   a  torus  of  molecular   gas  (the
Circumnuclear  Disk, CND).   The  region also  contains  three high-mass  star
clusters:  the Quintuplet,  the Arches,  and the  parsec-scale  cluster around
\SgrAstar.  This  region therefore presents very interesting  evidence of star
formation in this  peculiar part of a galaxy and  should help in understanding
starburst galaxies as well as high-mass star formation.

In the  early images  of the  central region of  the GC  recorded in  the near
infrared, at the best  seeing-limited resolution, several bright point sources
dominate the $\sim$\,20\arcsec$\times$20\arcsec\/  field centred on \SgrAstar. 
Among  these,  source GCIRS~16$\,$\footnote{The  sources  named ``GCIRS''  for
  Galactic  Centre Infrared  Source are  often referred  to simply  as ``IRS''
  sources in the GC-centric literature.}   was extremely bright and an intense
point source of  \lwl{\HeI}{2.058}.  This source has been  since then resolved
into a cluster of  six stars, but the same remarks still  apply to each of the
components: these six stars are very  bright, and exhibit intense \HeI\ lines. 
Stellar  classification of these  stars can  now be  attempted: they  are very
likely evolved OB  stars in a transitional phase  \citep{morris96}, very close
to  the Luminous  Blue  Variable (LBV)  stage.   Several dozens  of even  more
evolved  stars  (Wolf-Rayet stars)  have  been  observed  in the  same  region
\citep{krabbe95,paumard01}, and  massive stars are known to  orbit the central
black    mass    at    distances    as    short   as    a    few    light-days
\citep{schoedel03,ghez03,ghez04}.

These observations  show that massive star  formation has occurred  at or near
the GC within the last few million years.  Neither the mechanisms that lead to
massive star formation nor those that may lead to any star formation at all in
the vicinity of a SMBH are  currently known. Two basic types of scenarios have
been developed to  explain the presence of these stars  where they are: either
they  have been  formed  \emph{in situ},  or  they have  been  formed at  some
distance,  and  then  drifted  to  where  we see  them.  Both  hypotheses  are
problematic; they  will be  discussed in \sect{scenarios}.   In any  case, the
exact properties of these stars must  be studied in order to provide ground on
which to  build formation scenarios.   These properties include  exact stellar
type,  stellar  rotation, spatial  distribution,  radial  velocity and  proper
motion,  and require  both  high-resolution  imaging on  a  large enough  time
baseline and spectroscopy of each source to provide all this information.

Spectroscopic analysis with adaptive optics of GC sources has been performed a
number of  times using long  slits.  However, this  approach can lead  only to
limited results, because the data  contain information only for a very limited
number of sources, and because it is not possible to unambiguously associate a
given spectrum with a particular star. The second classical approach is to use
either  multi-band imaging  to  determine colour  indices  and thereby  colour
temperatures, or narrow band imaging  around spectral lines typical of certain
stellar classes. Proper  reduction of these imaging data  is however extremely
difficult because of the highly variable extinction \citep{blum96}, that makes
it quite hard to resolve the  degeneracy between temperature and \Av\ (we will
however  show  a  successful,  though   limited,  use  of  this  technique  in
\sect{IRS13}).  Concerning  narrow band techniques,  even using a  very nearby
continuum filter, it is necessary to  apply a different reddening to each star
in  order  to  avoid  false  detections.  As  the  non-homogeneous  extincting
material  is mingled  with the  stellar content,  even very  nearby  stars (in
projection)  can  be affected  by  a  different  reddening.  It  is  therefore
ultimately not  possible to create  a perfect extinction map.   Furthermore, a
number of spectral features are  complex, showing both emission and absorption
(for instance P~Cygni profiles), which can cancel each other.

For all these  reasons, it appears that significant progresses  can be made in
the field of stellar populations studies in the GC by means of spectro-imaging
techniques, which  allow one to  simultaneously obtain spectra  for \emph{all}
the stars  contained within a  field of view.   We have used  successively two
instruments  in  order  to  perform  such  a  study,  BEAR  and  SPIFFI.   The
characteristics of  both instruments will  be discussed in  \sect{instru}, the
reduction  techniques in  \sect{reduction}  and the  results  obtained in  the
Galactic Centre in \sect{results}.

\section{Instruments and observations}
\label{instru}
BEAR is  a prototype, made of  the CFHT Fourier  Transform Spectrometer (FTS),
coupled with  a NICMOS 3 camera. As  such, it is an  Imaging Fourier Transform
Spectrometer.  The  FTS provides  a very high  spectral resolution,  which can
indeed be  chosen at observation time  depending on the  target.  In practice,
resolutions up to $30,000$ have been obtained in this mode.  The field of view
is  set  by  the  original  design  of  the FTS,  which  was  not  built  with
spectro-imaging in mind,  and is thus limited to  $24\arcsec$.  This prototype
does not make  use of an adaptive optics system and  its spatial resolution is
therefore seeing-limited.  Finally, the bandwidth  of data is limited, for two
reasons: first, to limit both observing time and data size, and second because
the multiplex property of the FTS makes the S/N ratio lower when the bandwidth
increases.  We have obtained  several datasets concerning the Galactic Centre:
in \lwl{\HeI}{2.058}  at a spectral resolution  of \kms{52.9}, as  a mosaic of
three subfields; and  in \Brg\ (\microns{2.166}), at a  spectral resolution of
\kms{21.3}, as a  mosaic of two subfields.  These data  are presented in depth
in \citet{paumard04}. They cover most of a $40\arcsec\times40\arcsec$ field at
a   resolution  of   $\simeq0.5\arcsec$.    This  instrument   has  now   been
decommissioned  with  the closure  of  the  CFHT  infrared focus;  however,  a
specifically  designed instrument  following the  basic concept  of  BEAR with
adaptive  optics  would provide  a  very useful  observing  mode  for the  new
telescopes,  allowing for large  field, high  spectral and  spatial resolution
spectro-imaging.


SPIFFI \citep{thatte98,eisenhauer00,eisenhauer02}  is a near-infrared integral
field spectrograph  to be commissioned  as part of  SINFONI at the VLT  in June
2004.  It allows  observers to obtain simultaneously spectra  of $1024$ pixels
in  a $32\times32$  pixel  field-of-view.  In  conjunction  with the  adaptive
optics  system MACAO it  will be  possible to  perform spectroscopy  with slit
widths sampling the diffraction limit  of an 8m-class telescope. SPIFFI covers
the near-infrared wavelength range from \microns{1.1} to \microns{2.45} with a
moderate spectral  resolving power ranging  from $R=1000$ to $R=4000$,  and is
based on a reflective image slicer and a grating spectrometer.
K-band data  have been obtained  during a  test run on  2003 April $8/9$  as a
mosaic  of  about  two  dozen  subfields  covering  a  total  field  of  about
$10\arcsec\times10\arcsec$   centred   on    \SgrAstar\   at   an   excellent
seeing-limited resolution of  $\simeq0.25\arcsec$.  The spectral resolution is
\kms{85}.

\section{Data reduction}
\label{reduction}
\subsection{The ISM and its subtraction}
\begin{SCfigure}
\includegraphics[width=0.5\hsize]{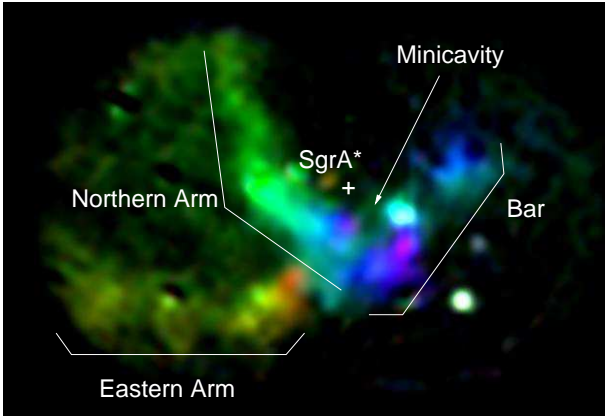}
\caption{The Minispiral as seen by BEAR in \Brg$\,^b$. This picture is a composite
  of three images  obtained from the BEAR cube through  three virtual filters. 
  The  colours give  therefore  the  velocity field  of  the Minispiral,  from
  \kms{-350} (red)  to \kms{+350}  (purple).  The standard  names of  some ISM
  features are given.\label{minispiral-brg} A  few emission line stars show up
  as point  sources: GCIRS~16NE  and 16C  just east of  \SgrAstar, The  AF and
  AF~NW stars below the Bar, and  the GCIRS~13E cluster on the western edge of
  the Minicavity.}
\end{SCfigure}
The first treatment  to find the spectral features is  to remove the continuum
emission from the  stars to obtain the \emph{line cubes}. This  can be done by
linear interpolation  between two neighbouring  continuum regions if  the cube
contains only  a narrow band (case of  BEAR) or by fitting  a simple function,
like a  polynomial, if the  continuum is not  nearly linear (case of  SPIFFI). 
Once the stellar continuum is subtracted, the BEAR \Brg\ data are dominated by
the       extended       emission       from       the       ionised       ISM
(Fig.~\ref{minispiral-brg}$\,$\footnote{The figures are available in colour at
  http://www.mpe.mpg.de/$\sim$paumard/YLU/   and   various  archives.}).    In
\citet{paumard04}, we have decomposed each  spectrum in the field into several
velocity components of the same line.  Assuming that the ISM is made of clouds
of finite velocity gradient material, we have decomposed the Minispiral into 9
components that are often superimposed on  each other.  In this paper, we have
shown  that these  features are  sufficiently thick  to be  responsible  for a
significant  extinction,  on  the order  of  half  of  the  K flux.   This  is
consistent with  the high  spatial variability of  \Av\ in this  region.  This
decomposition  required the very  high spectral  resolution provided  by BEAR,
because the lines  from two structures along the same line  of sight are often
very close to  one another in the spectral domain. In  some cases indeed, this
spectral separation goes down to  \kms{0}.  In these cases, interpolation must
be  used to  perform  the decomposition  using  information from  neighbouring
points where the two structures are sufficiently separated.

These ISM  features are  really intermingled with  the stellar content.  It is
therefore  necessary to  subtract the  interstellar contribution  in  order to
analyse   this   stellar   population.   In  \citet{paumard01},   updated   in
\citet{paumard03}, we  have again  used both the  high spectral  resolution of
BEAR  and its  spatial  properties in  the  \lwl{\HeI}{2.058} to  discriminate
between the stellar and interstellar lines, showing that some previous reports
of Helium stars  were indeed false, and due to  insufficient correction of the
extended emission.  To  clean the BEAR stellar spectra, we  simply cut out the
ISM lines  manually.  These  lines were identified  by visually  exploring the
cube and using the 2D information  it contained. This simple correction was of
good enough  quality because  at this very  high resolution, the  observed ISM
line width (\kms{\lesssim20}) was much  smaller than the spectral scale of the
stellar  features (\kms{\gtrsim150}).  For  these data,  we had  extracted the
spectra of previously  reported He-stars, and visually inspected  the cube for
more stars.  We hence reported three new detections in the field, out of which
one (N6) turned out to be essentially an unresolved thread of ionised ISM, the
two others being now confirmed by SPIFFI.

\label{ism-correction}
The SPIFFI data lack the spectral resolution required to clean each individual
spectrum of the narrow ISM lines it contains, but the spatial resolution is so
high (and it will still improve with the upcoming adaptive optics system) that
it is now  possible to reliably estimate the  extended line emission component
by interpolating each  frame of the cube over the locations  of the point line
emission or absorption  sources.  More specifically, the steps  that have been
applied for  this correction  are the following:  (1) determination of  the CO
index (a measure of the depth of the \microns{2.3} CO absorption band) of each
star using  narrow-band NACO  images; (2) the  stars with  a low CO  index are
suspected early-type  stars; (3)  for each spectral  channel of the  line cube
(i.e.   continuum  subtracted), an  aperture  corresponding  to each  possible
early-type star is marked as unavailable data; (4) each frame of the line cube
is interpolated  over these unavailable  regions; (5) a  low spatial-frequency
pass filter is  applied to each frame, the cube resulting  from these steps is
the ISM  cube, it  normally contains only  the ISM  emission plus most  of the
noise;  (6) this  ISM cube\label{ism-cube}  is then  subtracted from  both the
original  cube  and  the line  cube  to  obtain  the \emph{stellar  cube}  and
\emph{stellar  line  cube},  out  of  which  a  spectrum  for  each  suspected
early-type star is extracted.

\subsection{Identification of the featured stars}
This  leads  to  over  100  spectra,  most of  which  actually  show  spectral
signatures typical  of early-type  stars.  However, at  this point, it  is not
clear which  feature really belongs  to which star,  as the spectrum  of every
star  is  contaminated by  the  wings of  the  neighbouring  stars.  The  most
straightforward  way of resolving  this degeneracy  is visual  inspection: for
each feature of each spectrum, it is possible to extract from the stellar line
cube an image integrated exactly over the spectral domain corresponding to the
maximum emission (resp.   absorption) of the given feature.   In most cases, a
local maximum  (resp. minimum) appears  on this image  in the vicinity  of the
studied star.  The  exact location of this extremum reveals  to which star the
feature belongs, i.e., to the studied  star or to one of its neighbours.  This
visual method is inappropriate for heavy duty work such as that implied by the
analysis of the GC content,  which, as already mentioned, contains hundreds of
candidates and dozens of stars actually showing features accessible to SPIFFI.
In particular, this method  requires considerable interaction by the observer,
and  does not  make  use  of the  fact  that a  given  early-type star  should
typically show more  than one feature.  Furthermore, the  features do not have
to be  simple, the lines  often show P~Cygni  profiles, or consist  of several
lines from different species, both in emission and absorption (for instance, a
typical  feature  at \microns{2.11}  is  made of  two  \HeI\  lines, often  in
absorption,  one of  which  is itself  a doublet,  and  one N  line, often  in
emission  for O  stars). It  would therefore  be more  appropriate to  use the
entire spectra  to determine the spatial  location of the stars  to which they
belong.

\begin{figure}
\begin{minipage}{0.45\hsize}
\includegraphics[width=\hsize]{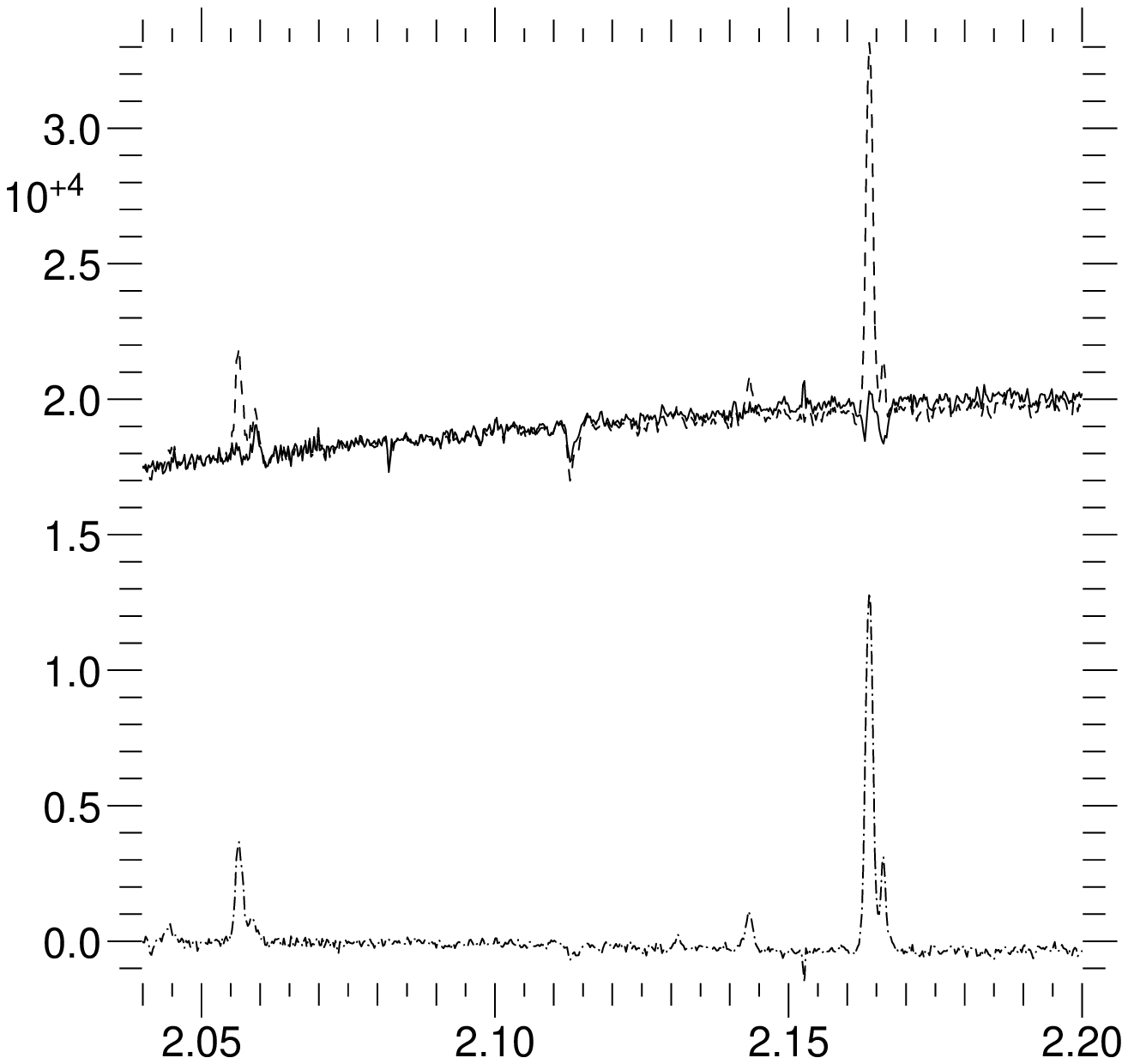}
\caption{Dashed spectrum: one of the original spectra (GCIRS~33N); 
  dash-dotted: the corresponding ISM spectrum; solid: the corrected stellar
  spectrum.\label{SPIFFI-spectrum-cleaning}}
\end{minipage}
\hfill
\begin{minipage}{0.45\hsize}
\includegraphics[width=\hsize]{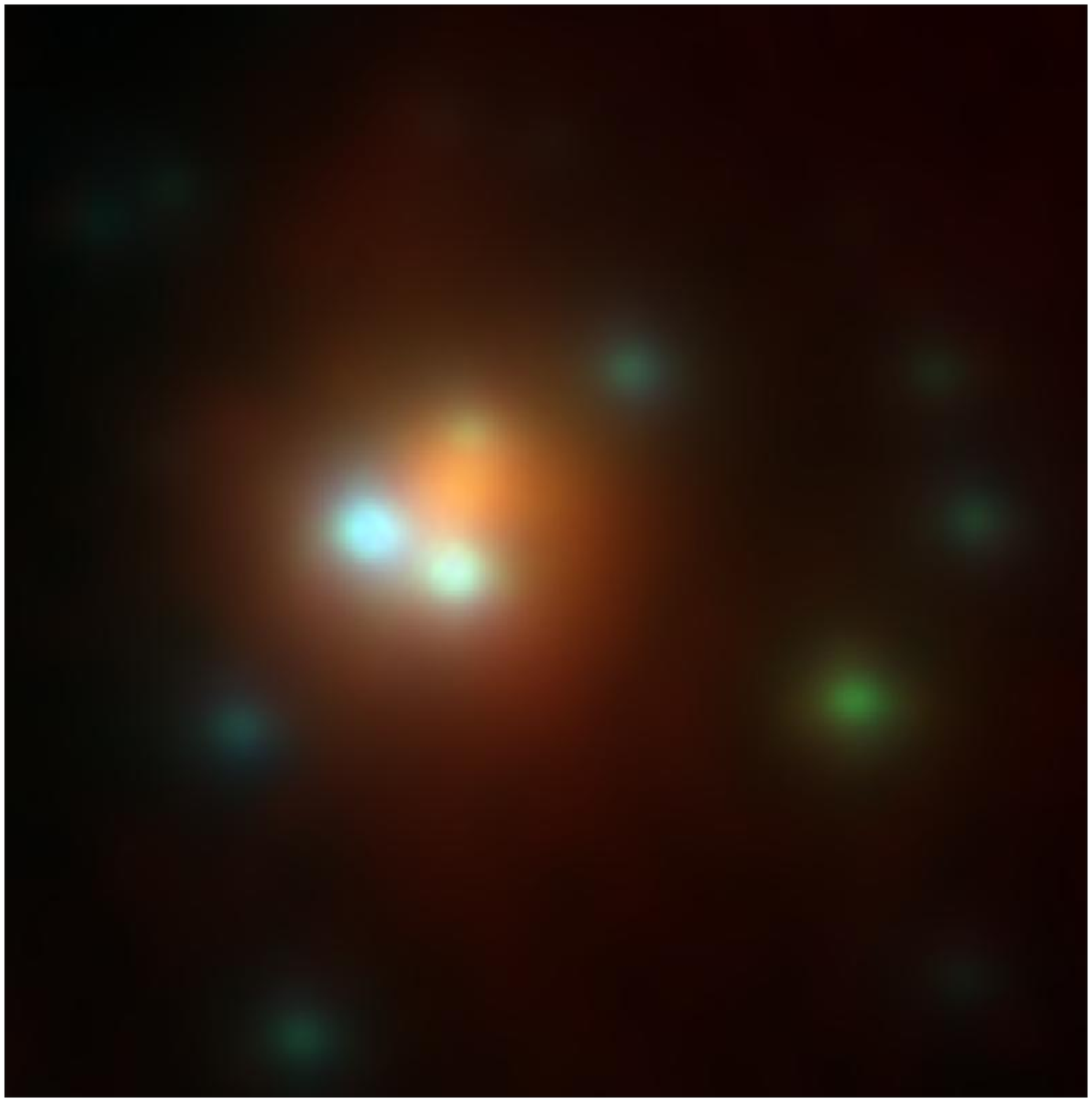}
\caption{HK'L composite image of GCIRS~13E$\,^b$.\label{IRS13-RGB}}
\end{minipage}\hfill
\end{figure}
We  have  developed a  method  specifically  to  achieve this  purpose.   Each
extracted stellar  spectrum can be  considered as a  template that we  want to
compare with  the individual spectra at each  pixel of the field,  in order to
determine the point  in the field from which this  spectrum arises and spreads
because of the spatial PSF.  This  comparison is done by means of correlation:
for  each  template spectrum  (each  spectrum  previously  extracted from  the
stellar line cube),  a correlation map is built.  This  map contains, for each
pixel  in the  field,  the correlation  factor  between the  template and  the
spectrum  contained  in the  stellar  line cube  at  this  location. This  map
normally contains  a local maximum at the  true location of the  star to which
the features  of the template spectrum  belong.  This technique  can easily be
automated,  and takes the  entire continuum-subtracted  spectrum of  the stars
into account.  Furthermore, this technique  can be generalised to provide both
a complementary detection  technique of the candidate early-type  stars, and a
first automated spectral classification of the stars.  In the method described
above, one can allow for a Doppler shift between the template spectrum and the
spectra  in  the  cube.   This  corresponds  to tracing  the  maximum  of  the
cross-correlation  function  for each  spatial  location  rather simply  using
correlation.  In that  case, every other star in the  field showing a spectrum
similar to the template  will show up as a local maximum  on the map.  It will
therefore be possible  to automatically determine which stars  have a spectrum
similar to the template.

When  this  cross-correlation work  has  been  done,  many candidates  can  be
rejected, as they  do not show detectable features.  Other  stars can be added
to   the  list   of  candidates   because  they   have  been   found   on  the
cross-correlation  map of  one  of the  templates.   The extraction  procedure
described above  can then be iterated  to take this information  into account. 
This is necessary  because information concerning the ISM  emission should not
be wasted  by blanking out the  region containing a  candidate early-type star
that  has been  shown  indeed not  to  show detectable  features;  and on  the
contrary the features  from the newly detected candidates  must be cleared out
as well as possible.  This second  extraction led to $29$ spectra, all of them
showing  recognisable features  typical  of early-type  stars, and  associated
unambiguously with a single star  (within the precision allowed by the spatial
resolution of the instrument).

All the  stars previously  studied with  BEAR within the  field of  SPIFFI are
detected here  as well.  As  already stated, object N6  from \citet{paumard01}
does not correspond to an He star but to a tiny ISM feature; however, this ISM
feature coincides in  projection, and may be physically  associated, with four
stars that indeed  show some He feature (ID 24, 26,  31 and 33).  Furthermore,
for the  first time, thanks to their  high spatial resolution as  well as high
signal-to-noise  ratio, these SPIFFI  data allow  detection of  faint emission
lines as well as absorption lines, which are totally filled by ISM emission on
the raw  data.  Fig.~\ref{SPIFFI-spectrum-cleaning} shows one  of the original
spectra, the  ISM profile determined  as in \sect{ism-cube}, and  the inferred
stellar spectrum.   After correction for  the very intense  interstellar \Brg\ 
line, a complex stellar feature made of \Brg\ and \HeI\ in absorption appears.

\section{GCIRS~13E: a case for spectro-imaging}
\label{IRS13}
\begin{figure}
\includegraphics[width=0.3\hsize]{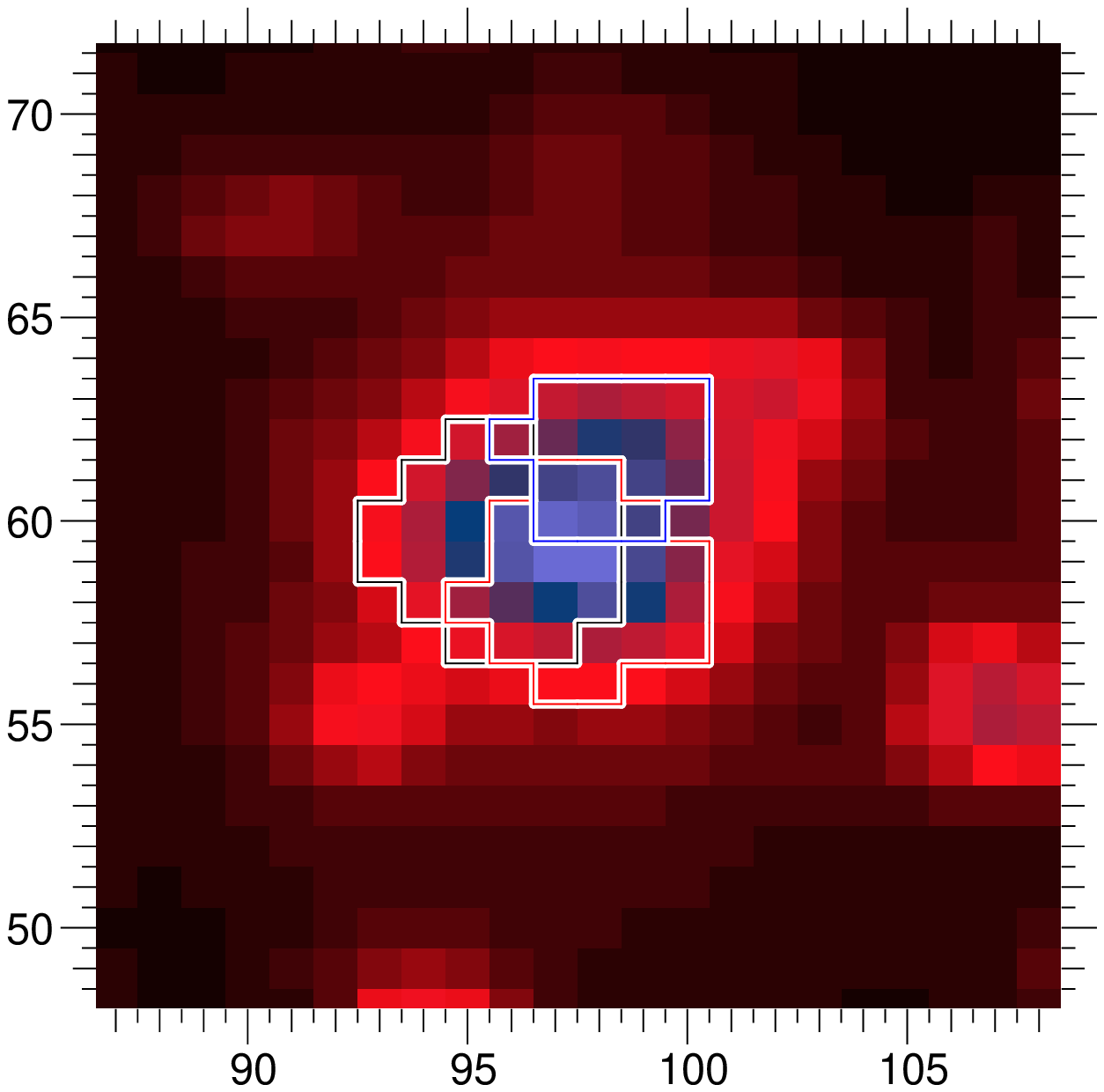}a)\hfill%
\includegraphics[width=0.3\hsize]{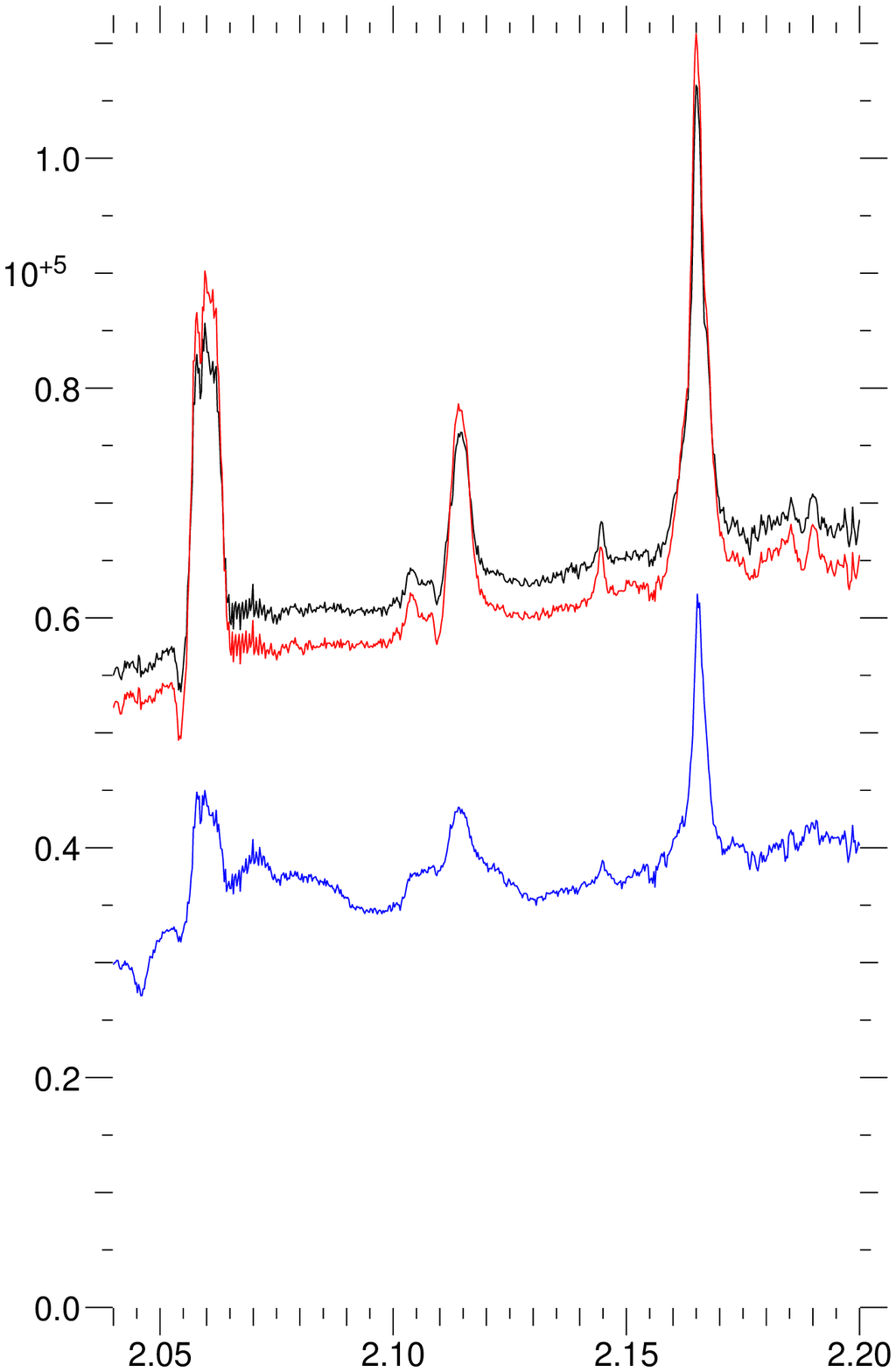}~b)\hfill%
\includegraphics[width=0.3\hsize]{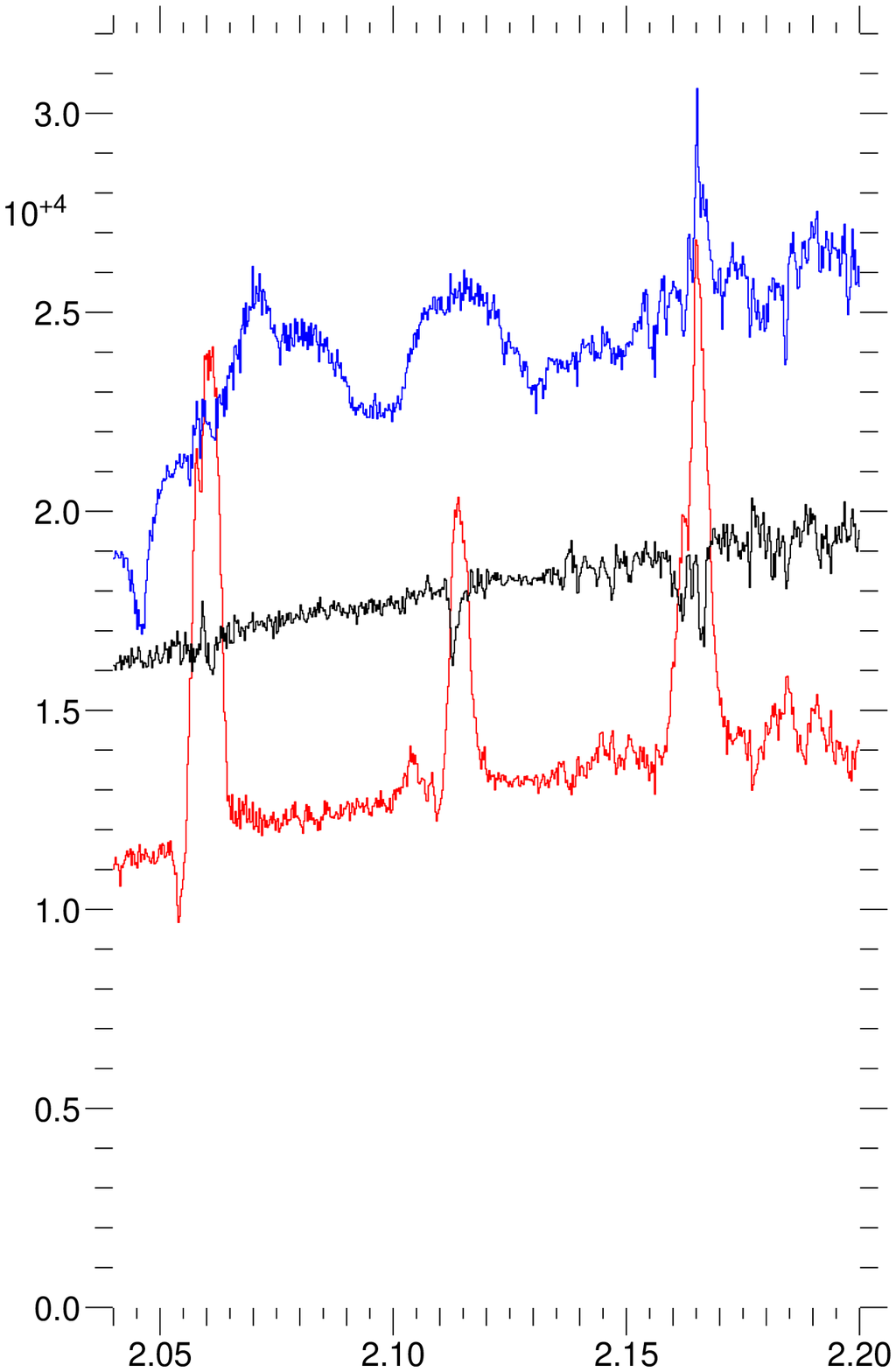}~c)
\caption{SPIFFI spectra of GCIRS~13E: \textbf{a)} apertures used,
  \textbf{b)}    raw    (blended)    spectra,   \textbf{c)}    final    object
  spectra.\label{IRS13-aper-spectra}}
\end{figure}
Among  these objects, GCIRS~13E  deserves special  attention.  This  source is
bright at all wavelength from sub-mm to X rays. A three colour high-resolution
image (Fig.~\ref{IRS13-RGB})  made from Gemini  H and K'  and ESO 3.6m  L band
data  illustrates that  it contains  three blue  stars and  a very  red  core. 
Deconvolution  of   these  three  images   plus  narrow  band   NICMOS  images
\citep{maillard04} has  shown that GCIRS~13E is  indeed a cluster  of at least
seven evolved  massive stars: even the  three red objects at  the cluster core
can be  interpreted as dusty  Wolf-Rayets.  \Av\ can  be derived from  the two
assumptions that  it does not vary  significantly within the  cluster and that
the      bluest     star     is      in     the      Rayleigh-Jeans     regime
($T_\text{eff}\gtrsim25\,000$~K at  this wavelength),  which is proven  by the
fact  that two  of the  stars show  emission in  a NICMOS  \Paa\  image.  This
multi-wavelength, high spatial resolution study can be considered as a kind of
low spectral  resolution attempt.  However,  the exact spectral type  of these
hot stars cannot be unambiguously  established by this technique, because more
spectroscopic  information is  needed.  Therefore,  spectra of  the individual
components are required to achieve this purpose.

The sources  are too  close to  one another to  allow for  individual aperture
spectroscopy. Any attempt  to do this is doomed  to give spectra significantly
affected  by one  another.   In Fig.~\ref{IRS13-aper-spectra},  we show  three
spectra  obtained  from the  SPIFFI  data  through  three overlapping  virtual
apertures; two of  them seem identical, whereas the  third is clearly affected
by  a   blend  with  these.    Reducing  these  apertures  would   reduce  the
signal-to-noise ratio  while not providing much better  results.  However, the
spectrum obtained through each of the apertures is a linear combination of the
spectra of all the stars (at least seven indeed).  It is therefore possible to
obtain three independent spectra by linear combination of these three observed
spectra,  leading to spectra  typical of  three of  the types  discussed below
(Fig.~\ref{IRS13-aper-spectra}c),  although it  is not  yet clear  whether the
upper spectrum in this figure belongs to the blue star north of the cluster or
to the red stars in the middle.

\section{The (known) stellar population of early-type stars}
\label{results}
\label{BEAR-results}
\begin{figure}
\begin{minipage}{0.48\hsize}
\includegraphics[width=\hsize,bb=74 386 603 899]{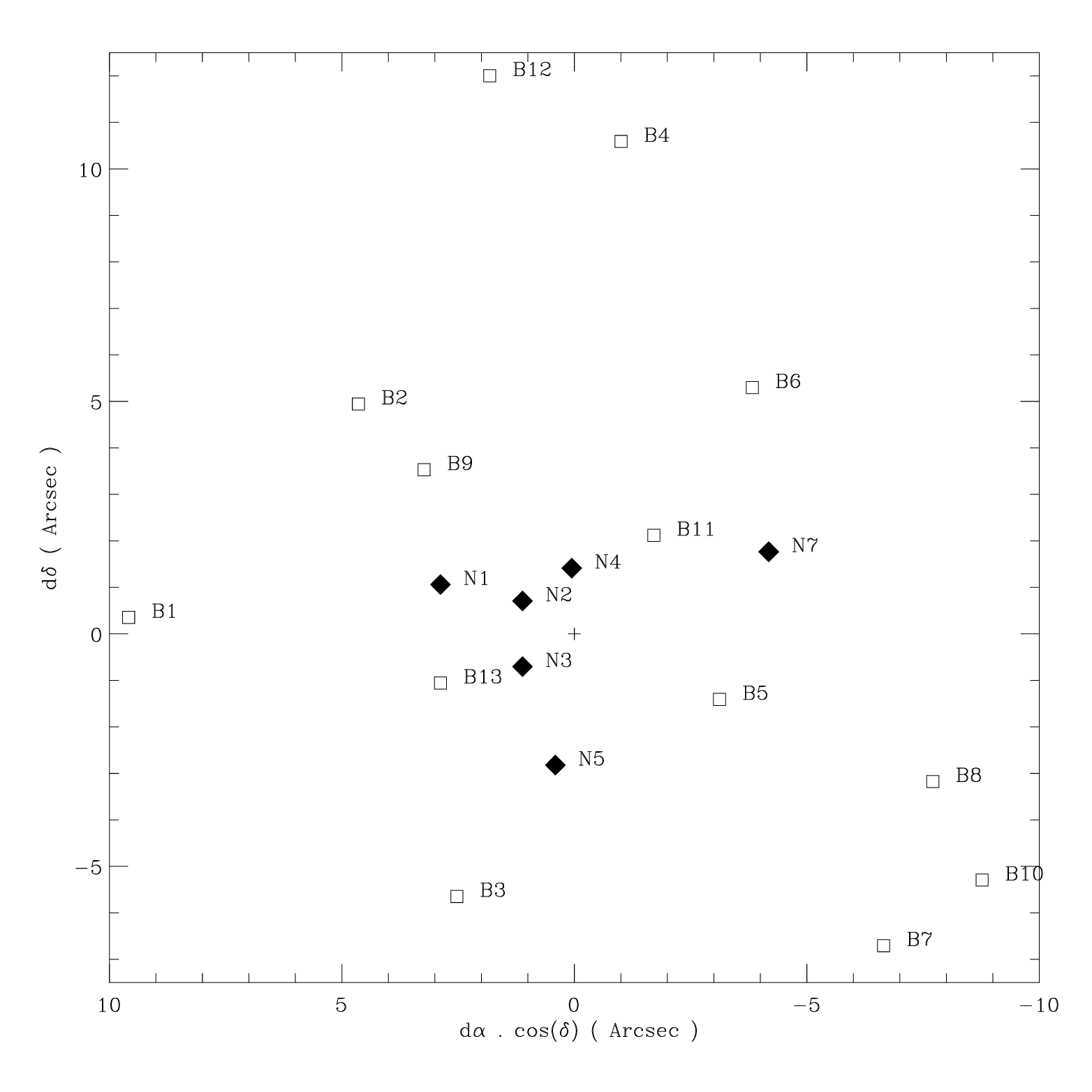}
\caption{Spatial distribution of the narrow (N) and broad (B) line stars.\label{star-map-BEAR}}
\end{minipage}\hfill
\begin{minipage}{0.48\hsize}
\includegraphics[width=\hsize]{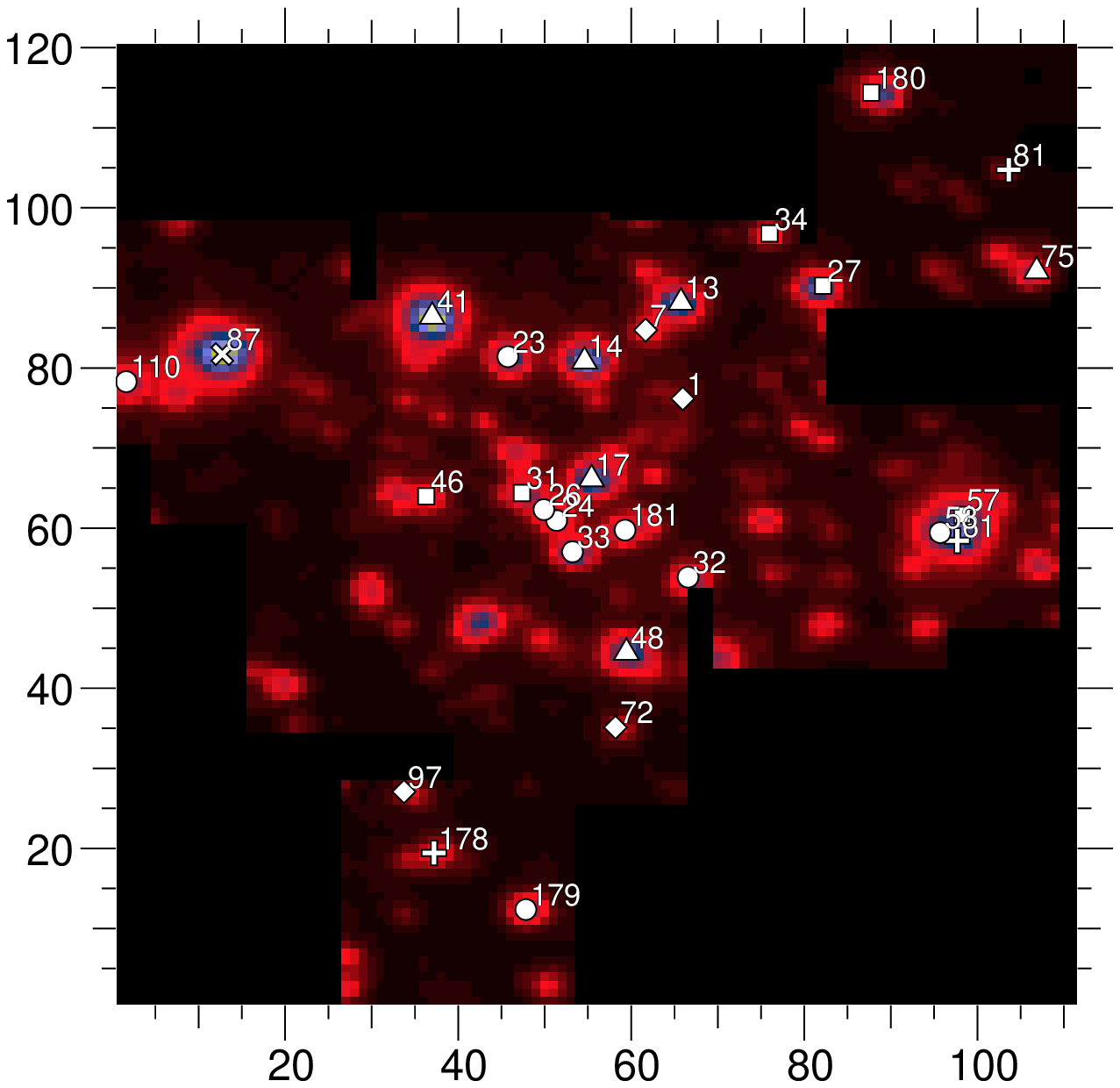}
\caption{Spatial distribution of the 6 stellar types:
  early  (squares)   and  late  (vertical  crosses)   Wolf-Rayet  stars,  LBVs
  (triangles), OBN stars (circles), a  candidate Be star (diagonal cross), and
  a few more OB stars.\label{SPIFFI-types-map}}
\end{minipage}
\end{figure}

\begin{figure}
\includegraphics[width=0.3\hsize]{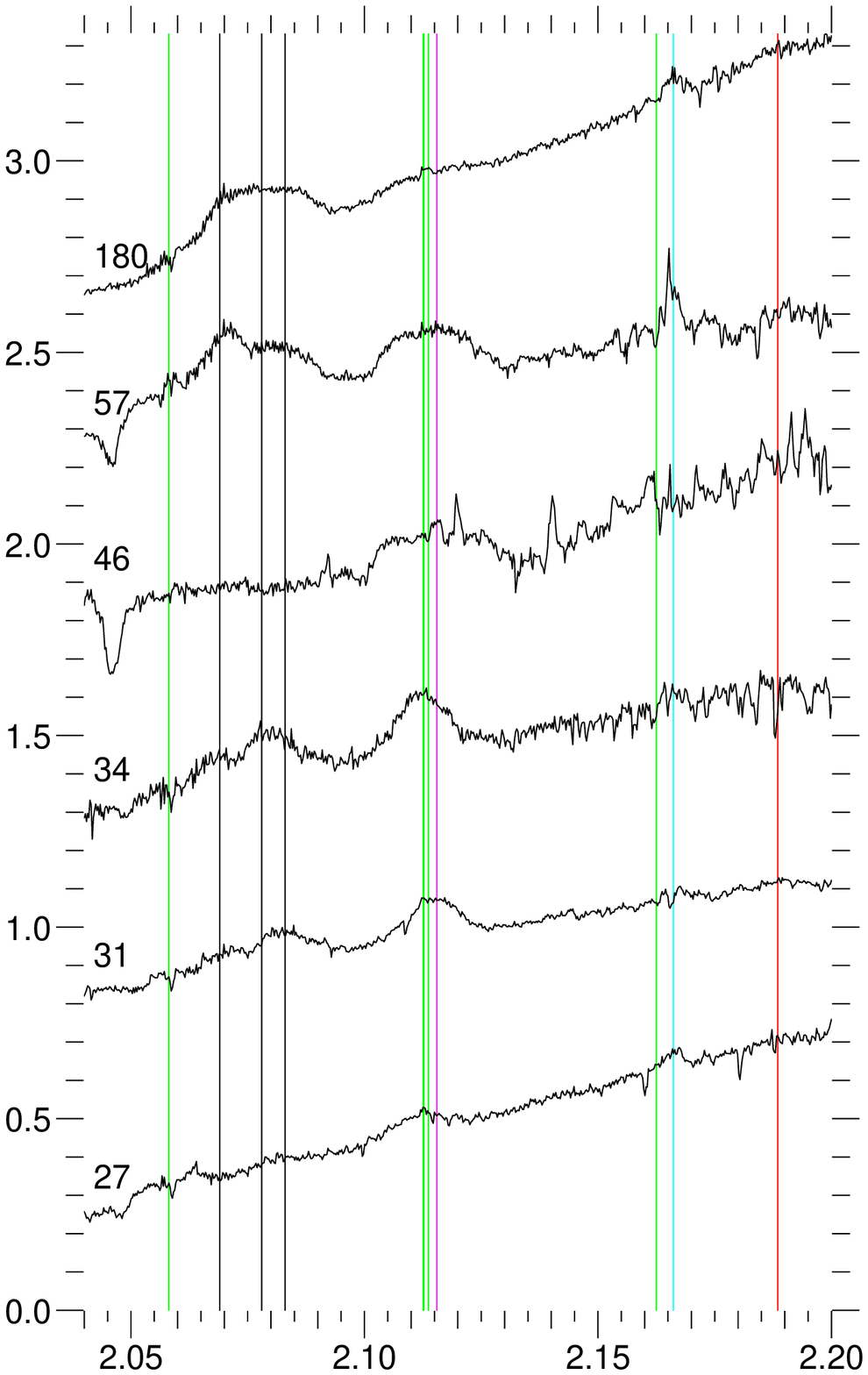}\hfill
\includegraphics[width=0.3\hsize]{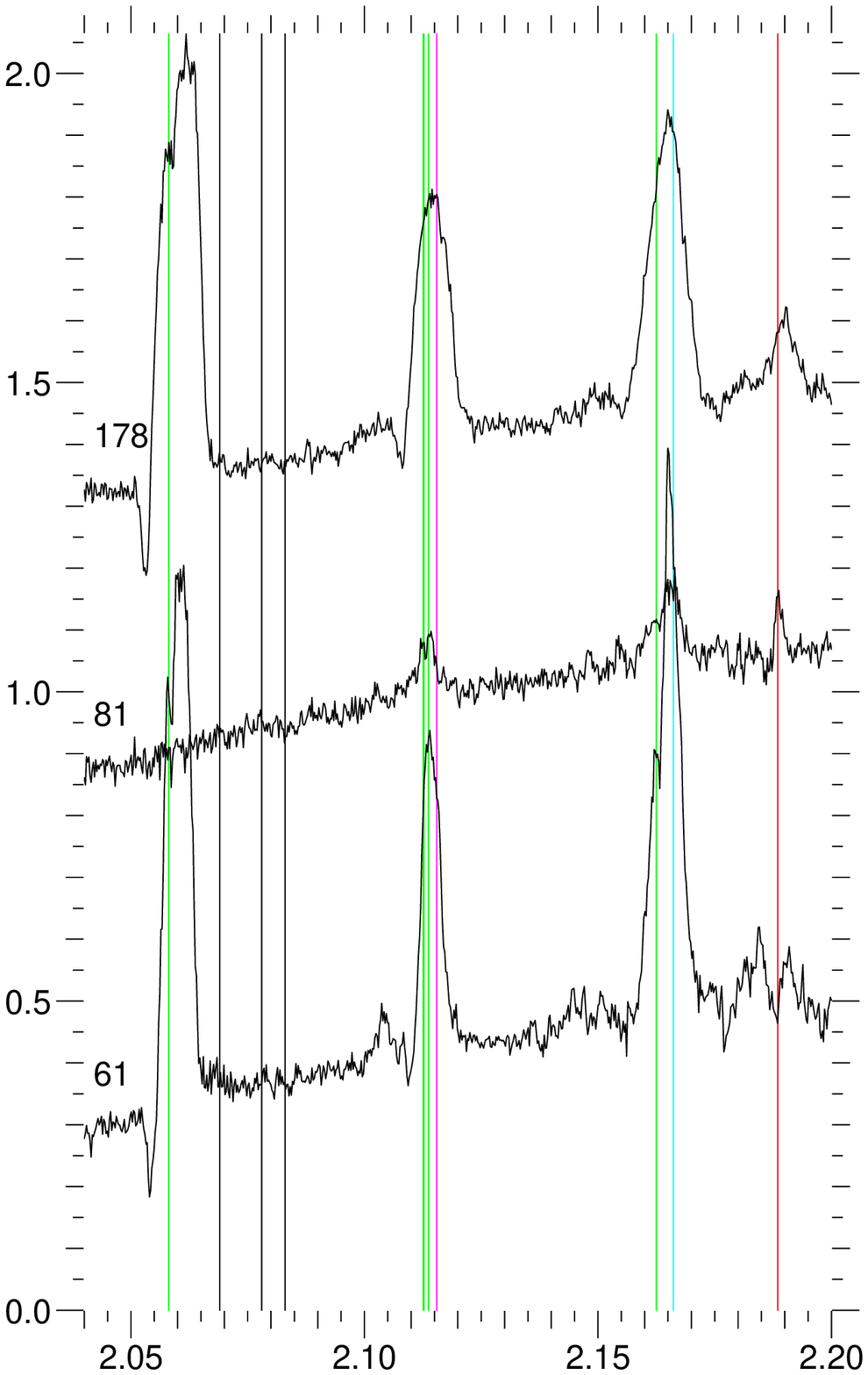}\hfill
\includegraphics[width=0.3\hsize]{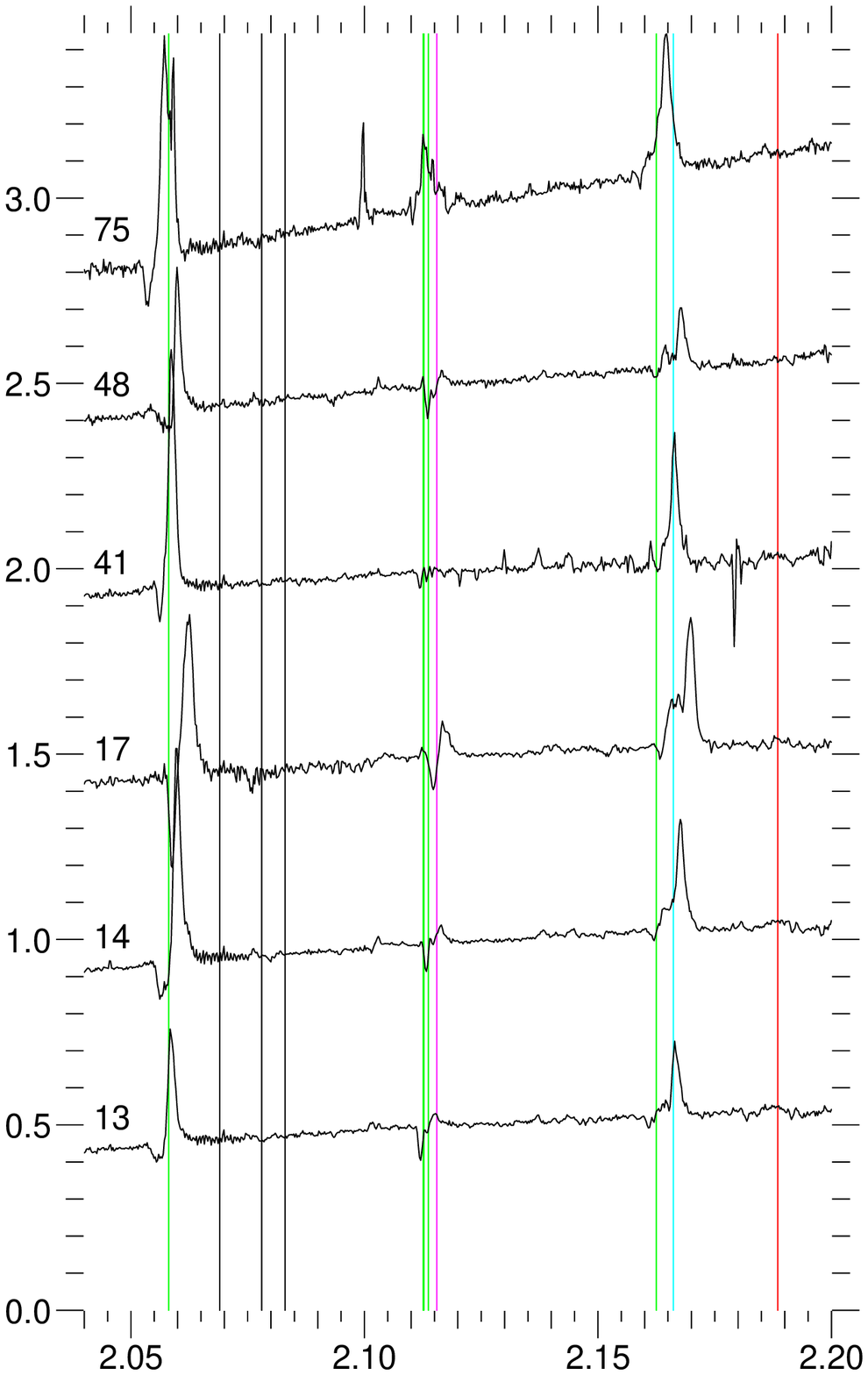}\\
\begin{minipage}{0.3\hsize}
\caption{Early Wolf-Rayets.\label{SPIFFI-spectra-1}}
\end{minipage}
\hfill
\begin{minipage}{0.3\hsize}
\caption{Late Wolf-Rayets.\label{SPIFFI-spectra-2}}
\end{minipage}
\hfill
\begin{minipage}{0.33\hsize}
\caption{Luminous Blue Variables.\label{SPIFFI-spectra-3}}
\end{minipage}
\end{figure}

\begin{figure}
\begin{minipage}{0.3\hsize}
\includegraphics[width=\hsize]{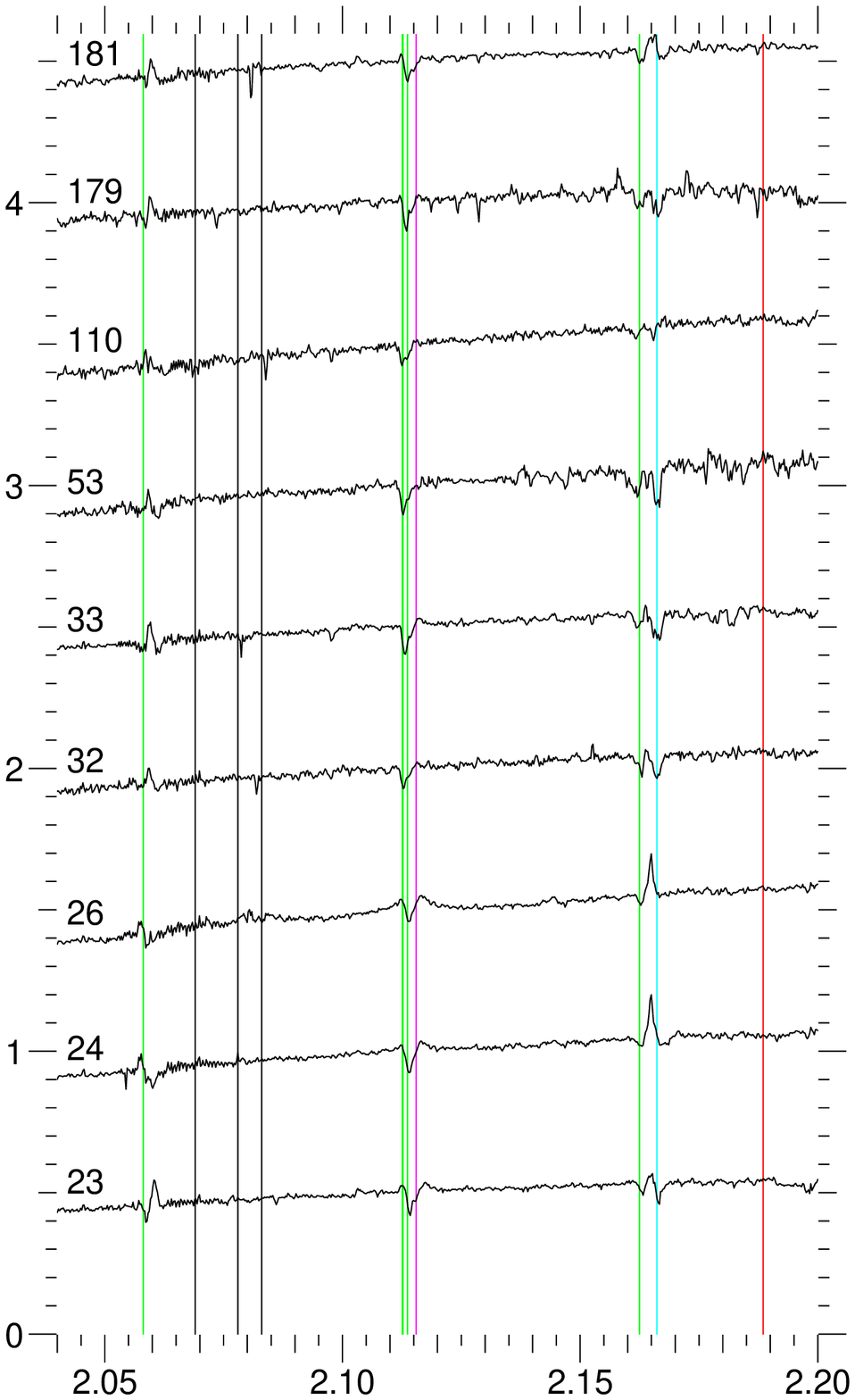}
\caption{N- and He-rich stars (OBN).\label{SPIFFI-spectra-4}}
\end{minipage}
\hfill
\begin{minipage}{0.3\hsize}
\includegraphics[width=\hsize]{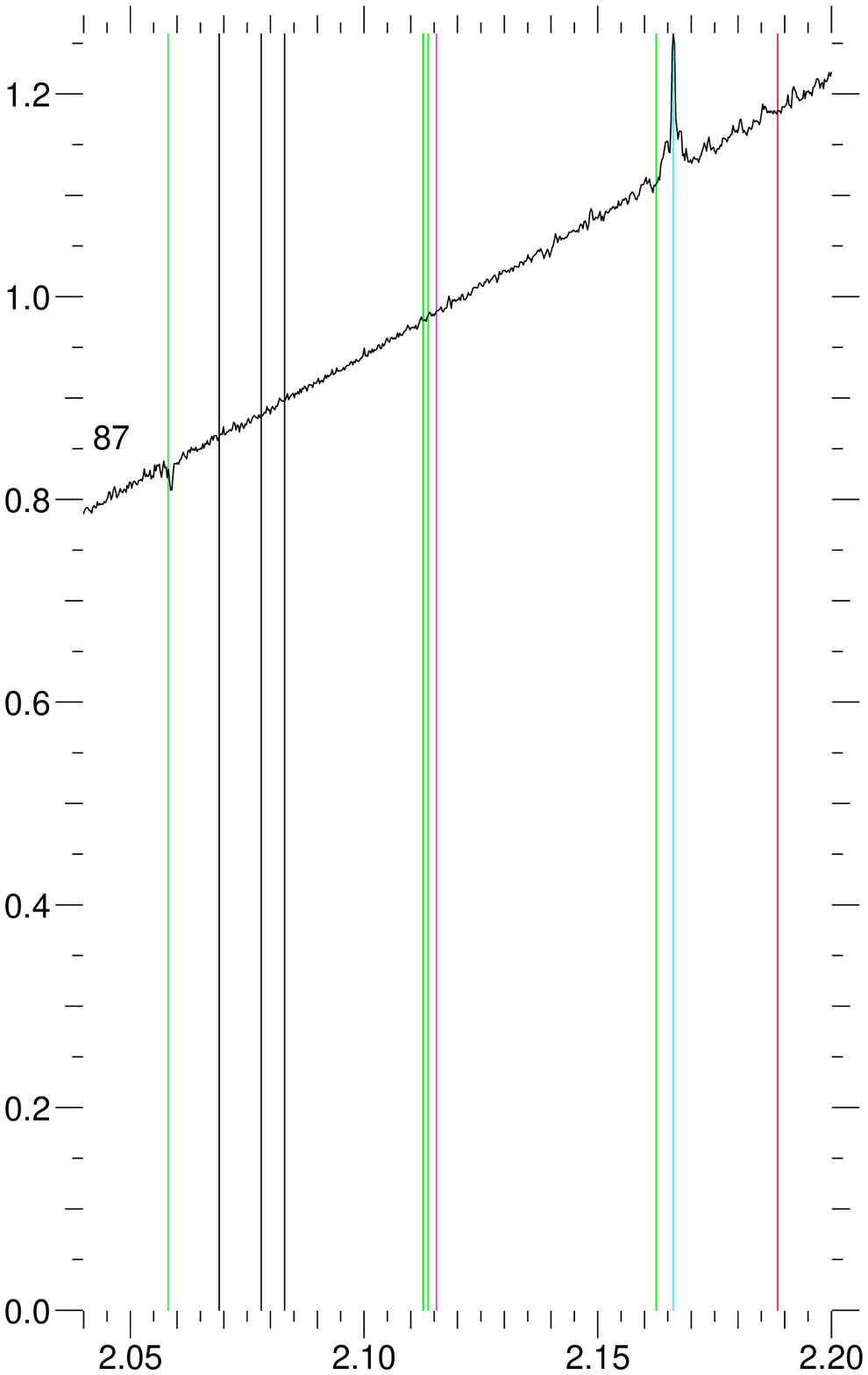}
\caption{GCIRS~1W, a candidate Be star.\label{SPIFFI-spectra-6}}
\end{minipage}
\hfill
\begin{minipage}{0.3\hsize}
\includegraphics[width=\hsize]{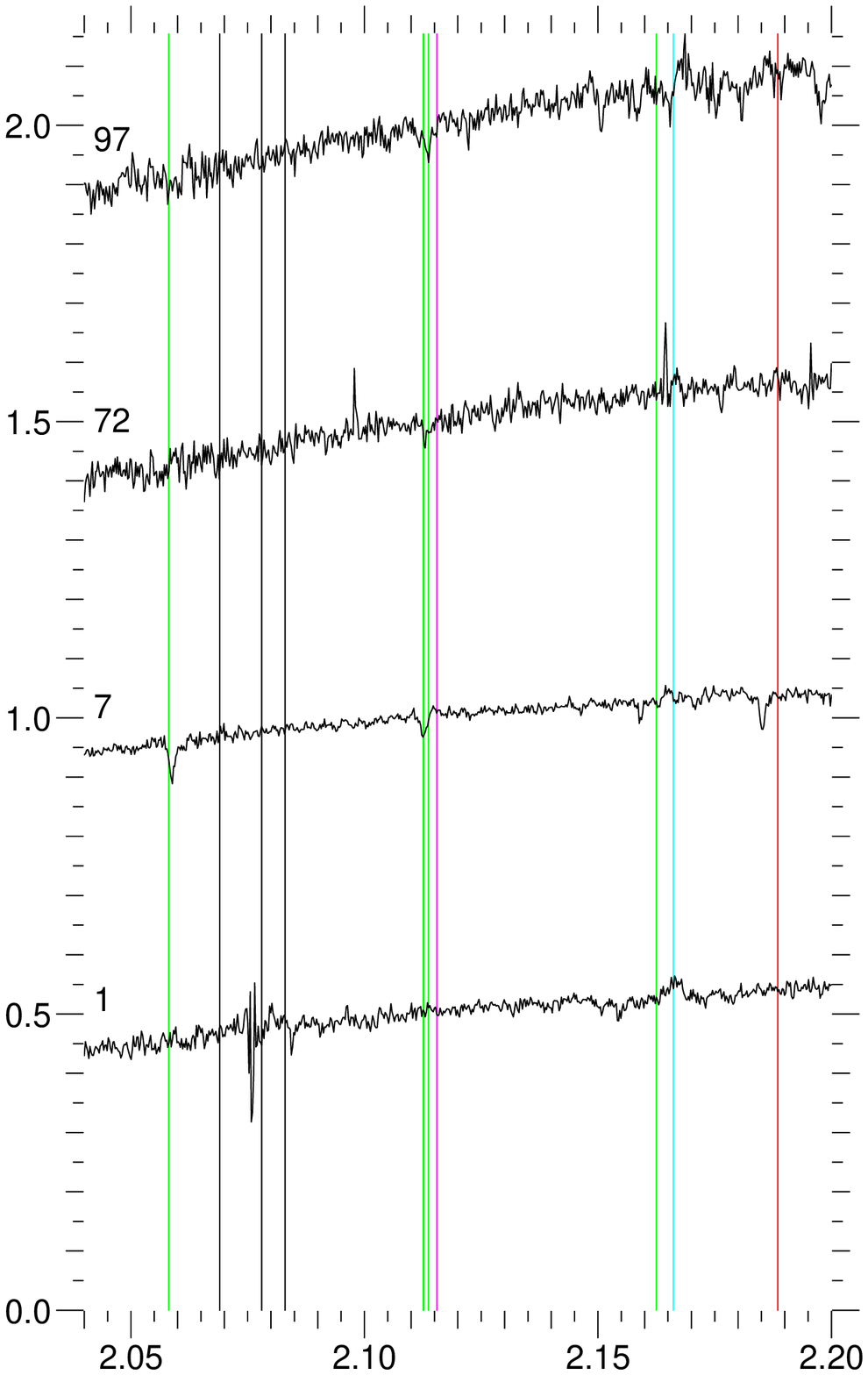}
\caption{More OB stars.\label{SPIFFI-spectra-5}}
\end{minipage}
\end{figure}

\begin{table}
\begin{minipage}{\hsize}
\caption{Translation of ID numbers in the figures to common names.
  Offsets from \SgrAstar\ (RA, dec. in  arcsec) are given for so far anonymous
  stars.\label{star-names}}
\begin{tabular}{lrl}
\hline\hline
Fig.~\ref{star-map-BEAR} & Fig.~\ref{SPIFFI-types-map} & Name \\
\hline
N1  & 41  & {IRS 16NE}  \\
N2  & 14  & {IRS 16C}   \\
N3  & 17  & {IRS 16SW}  \\
N4  & 13  & {IRS 16NW}  \\
N5  & 48  & {IRS 33SE}  \\
N7  & 75  & {IRS 34W}   \\
B1  &     & {ID~180}$^a$  \\
B2  &     & {IRS 7E2}   \\
B3  & 178 & {IRS 9W}    \\
B4  &     & {IRS 15SW}  \\
B5  & 61  & {IRS 13E2}  \\
B6  &     & {IRS 7W}    \\
B7  &     & {AF star}     \\  
\hline
\end{tabular}\hfill
\begin{tabular}{lrl}
\hline\hline
Fig.~\ref{star-map-BEAR} & Fig.~\ref{SPIFFI-types-map} & Name \\
\hline
B8  &     & {AF NW}       \\
B9  &     & {HeIN3}       \\ 
B10 &     & {BSD WC9}     \\  
B11 & 27  & {IRS 29N}   \\
B12 &     & {IRS 15NE} \\
B13 & 46  & {IRS 16SE2}\\
    & 1   & S2 (S0-2)    \\
    & 7   & S1-3 \\
    & 23  & IRS~16CC\\
    & 24  & $1.447$, $-1.49$\\
    & 26  & $1.593$, $-1.355$\\
    & 31  & IRS~16SE1\\
    & 32  & IRS~33N\\
\hline
\end{tabular}\hfill
\begin{tabular}{rl}
\hline\hline
Fig.~\ref{SPIFFI-types-map} & Name \\
\hline
     33  & MPE 1.6-6.8\\
     34  & IRS~29NE1\\
     53  & IRS~13E1\\
     57  & IRS~13E north$^b$\\
     72  & $0.774$, $-4.047$\\
     81  & IRS~34NW\\
     87  & IRS~1W  \\
     97  & $3.195$, $-4.842$\\
     110 & $6.372$, $0.227$\\
     179 & $1.8$, $-6.3$\\
     180 & IRS~3$^c$\\
     181 & $0.665$, $-1.608$ \\
\null&\null\\
\hline
\end{tabular}\\
{\footnotesize $a$: in \citet{ott99}; $b$: north of the cluster, exact identification 
uncertain; $c$: the emission line star is offset by about $0.1\arcsec$ 
from the bright K-band source.}
\end{minipage}
\end{table}
The main  results obtained with BEAR \citep{paumard01,paumard03}  are that the
He stars can  indeed be classified in two groups, from  their line width (mean
$\text{FWHM}\simeq\kms{225\pm75}$    for   the    narrow   line    stars   and
\kms{1025\pm400} for  the broad  line stars) and  K magnitude: all  the narrow
line  stars are more  luminous than  the broad  line stars,  by more  than $2$
magnitudes  in average (in  \citealt{paumard01}, we  reported that  one narrow
line star, GCIRS~34W, was less luminous than the others, but that was due to a
temporary obscuration  event, as discussed in \citealt{paumard03}  and below). 
A  third, striking  property of  these two  stellar classes  is  their spatial
distribution   (Fig.~\ref{star-map-BEAR}):   the   narrow   line   stars   are
concentrated  in  the GCIRS~16  complex,  whereas  the  broad line  stars  are
distributed in the entire  field. Table~\ref{star-names} gives the translation
between  the  identification  numbers  used  in  Fig.~\ref{star-map-BEAR},  in
Figs.~\ref{SPIFFI-types-map} to \ref{SPIFFI-spectra-5}, and common names.

The SPIFFI  data set that  we have  analysed so far  has a smaller  field, and
therefore  does  not add  much  information  on  these two  different  spatial
distributions.   However,  thanks   to  their  other  characteristics  already
discussed  (wide  band,  high  spatial resolution,  and  high  signal-to-noise
ratio), they give a lot of  interesting results. We will hereafter present the
spectra  of all  the early-type  stars  that we  have found,  together with  a
discussion of their  data type. The spectra presented here  are limited to the
range \microns{2.04-2.20},  for comparison with  \citet{conti}.  The reduction
software  being still under  heavy development,  some artifacts  remain, which
take  the   form  of   strong  noise  spikes.    The  spectra  are   given  in
Figs.~\ref{SPIFFI-spectra-1}  to  \ref{SPIFFI-spectra-5},  and  their  spatial
distribution  is given in  Fig.~\ref{SPIFFI-types-map}.  The  various vertical
lines on  Figs.~\ref{SPIFFI-spectra-1} to \ref{SPIFFI-spectra-5}  mark several
atomic lines: green: \HeI, pink: carbon, red: H (\Brg), blue: \HeII.

\textbf{Early Wolf-Rayet stars:}
Six stars show very broad features  mainly from \ion{C}{iv} and/or N; they are
early  Wolf-Rayet stars  (WC and  WNE).   Some of  them also  show some  \Brg\ 
emission.  GCIRS~16SE2 had been mistaken  for a broad \HeI\ emission line star
in \citet{paumard03}. This was due to the fact that an unidentified absorption
feature  at roughly  \microns{2.045} lays  in our  short  wavelength continuum
region, which  mimicked a low signal-to-noise,  broad emission line  on a very
red continuum. GCIRS~29N  shows broad and tenuous features  of \ion{C}{iv} and
N, but  also \HeI; it was  the broad \HeI-line stars  B11 in \citet{paumard03}
and may be  of a slightly later type  than the other stars of  this group. Its
weak \lwl{\HeI}{2.058} resembles that of  the star Blum~WC9 (B10), which could
therefore be of the same type.  These seven stars are randomly spread over the
entire field.

\textbf{Late Wolf-Rayets:}
The second  subset of stars is  typical of late Wolf-Rayet  stars (WNLs), with
rather broad  and bright \HeI, \HeII,  \Brg\ and N  lines. Two of them  show a
broad \lwl{\HeI}{2.058} line, typical  of the broad-line He-stars discussed in
\sect{BEAR-results}.   All these  broad-line stars  (except B10,  B11  and B13
already discussed) therefore seem to be WNLs as well.

\textbf{Luminous Blue Variables:}
The third  subset corresponds  exactly to the  narrow-line stars  discussed in
\sect{BEAR-results}.   In this  wavelength domain,  they are  characterised by
their rather strong, narrow \lwl{\HeI}{2.058} line with a clear P~Cyg profile;
an equally strong feature at \microns{2.166} made mainly of \Brg\ as well as a
clear contribution  of \lwl{\HeI}{2.163}; this  complex does not show  a P~Cyg
profile, but  that may be mainly because  of the blending; a  complex of \HeI\ 
(in  absorption   for  all   but  one  star)   and  N  (in   emission)  around
\microns{2.11}.
These  spectra are  typical of  so called  ``transitory'' stars,  Ofpe/WN9 and
Luminous Blue Variables (LBVs).  \citet{morris96} seems to show that all these
types may relate to objects of the same nature, maybe in different states.  As
already stated  these stars  are all within  the GCIRS~16  cluster.  GCIRS~34W
deserves special  attention: this  is the  only one to  have already  shown an
obscuration event,  which fully qualifies it as  an LBV; this is  the only one
that shows the \lwl{\HeI}{2.11} complex in  emission; and this is the star for
which  the  spatial  association  with   the  GCIRS~16  cluster  is  the  most
controversial.
  The first point above does not mean that the other stars are not LBVs as well,
as  obscuration events  of  LBV  stars are  indeed  rare \citep{humphreys99}.  
Neither does the second, as the spectra  of LBVs are variable as well as their
luminosity:  the  fact  that  the  \lwl{\HeI}{2.11} complex  of  GCIRS~34W  is
currently in emission could be related to its obscuration event.

\textbf{OBN stars:}
The third type is defined by stars which show: \lwl{\HeI}{2.058} (often with a
complex  structure);   the  \lwl{\HeI}{2.11}  complex   in  absorption;  often
\lwl{N}{2.11} in  emission; comparable \Brg\ and  \lwl{\HeI}{2.163}, mostly in
absorption, but sometimes with some  emission in-between.  Note that the three
stars that show emission at \microns{2.166}  are in a crowded area, and two of
them are  superimposed upon a  very compact thread  of ISM material.   The ISM
correction at  this very  location is therefore  somewhat suspicious.   We had
reported a He star at this location with BEAR, but the same comment applies to
this detection as well.

Following the  independent stellar classification scheme in  the K-band (hence
the ``k'' prefix)  by \citet{conti}, these stars can  be classified in various
kOB subtypes (kOBh+, kOBh-, kO7-O8, kO9-B1b+,  kOBb and kOBbh+). It is hard to
give an unambiguous  classification in the MK system for  these stars, for two
reasons:  a unambiguous  relationship between  the MK  system and  the  K band
classification scheme  by \citet{conti} does fundamentally not  seem to exist;
and very few standard  stars have been studied so far both  in the optical and
infrared to  provide ground for comparison.  This is particularly  true for OB
stars.

However,  when one compares  directly these  stars to  the templates  shown in
\citet{conti},  it appears  clearly  that  only two  stars  exhibit a  similar
feature at \microns{2.166}, with comparable \Brg\ and \HeI: HD~191781 (kOBbh+)
and  HD~123008 (kO7-O8bh-).   Both are  ON9.7  supergiants.  ON  and BN  stars
(generically called  OBN stars) are  particular kinds of  O and B  stars which
show unusual  N lines in the optical.   They are also known  to show unusually
strong He lines, and to be  particularly bright because of a lower atmospheric
opacity \citep{langer92}.  Indeed,  our 9 stars are particularly  bright, as a
typical O9  supergiant should be  at most 1.4  magnitudes brighter than  S2 (a
O8-B0V star  according to \citealt{ghez03},  of $m_K=13.95$), whereas  their K
magnitudes span the 10.43--12.63 range.   It therefore seems likely that these
nine stars are indeed N- and He-rich stars that would appear in the optical as
OBN stars.

\textbf{GCIRS~1W: a Be star?}
GCIRS~1W is known to exhibit a very flat and very red spectrum, to be embedded
within  the  Northern  Arm and  to  interact  with  it  to  form a  bow  shock
\citep{tanner04}. Study of this star is  made difficult by the fact that it is
a local source  of excitation for the ISM, and is  therefore coincident with a
local maximum of  the ionised gas emission. However,  the correction described
in   \sect{ism-correction}   allows   us   to   unambiguously   identify   the
\lwl{\HeI}{2.06} line  in absorption as  well as the  \Brg\ line in  emission. 
This spectrum is similar to that  of the Be star DM~+49~3718 in \citet{conti}. 
Indeed,  all the  stars in  this atlas  which show  \Brg\ in  emission  and no
\lwl{\HeI}{2.11} are Oe or Be stars.

\textbf{More OB stars:}
Finally, a few more  stars exhibit at least one \HeI\ line  or the \Brg\ line. 
Star number 1 (S2) shows only to a detectable level the \Brg\ line, and we are
mostly able to detect it because  a very high Doppler shift (about \kms{1400})
puts it away  from the ISM residuals, at about  \microns{2.155}. This star and
others of  its kind will be more  easily studied when SPIFFI  is equipped with
adaptive optics. The three other stars show mostly the \lwl{\HeI}{2.163} line.
The absence of detection of \Brg\ is certainly due to the noise in that region
of the spectrum, which comes from the residuals of the ISM feature.

\section{Formation scenarios}
\label{scenarios}
The existence of the early-type  stars at their current position can basically
be explained in two ways: either they have been formed \emph{in situ}, or they
have been  formed somewhere  else in  the Galaxy before  they have  been drawn
here. Various items  in the literature argue that  gravitational collapse of a
gas cloud  is prevented in  the tidal field  of \SgrAstar, and  that therefore
star formation cannot have  occurred \emph{in situ}.  However, \citet{levin03}
show that stellar  formation in the vicinity  of a SMBH is likely  to occur if
the black hole  possesses a massive enough accretion  disk. This consideration
transforms the  problem into: did  \SgrAstar\ possess a massive,  parsec scale
accretion disk $10^6$ years ago?  The  fact is that \SgrAstar\ does not appear
to possess a  disk, of any observable  scale, right now.  The ISM  that can be
seen at the  parsec scale is concentrated in the CND  (which may contain about
\solarmasses{10^4}   of   material,   \citealt{christopher03})  and   in   the
Minispiral, for which \solarmasses{10^3} seems to be a reasonable upper limit,
as  the  estimates for  two  of  the nine  components  of  the Minispiral  are
\solarmasses{\simeq10}  \citep{liszt,  paumard04}.  However,  \citet{morris99}
show that  current absence  of a disk  remains compatible with  \emph{in situ}
formation within an accretion disk in the context of a limit cycle.

The other possibility  is that these stars have been  formed at some distance,
and  then moved  into the  central  parsec.  The  presence of  the Arches  and
Quintuplet clusters  at a  distance of about  $35$~pc shows that  massive star
cluster  formation  is  indeed  possible  at these  distances.   According  to
\citet{mcmillan03},     a    cluster    with     a    dense     enough    core
(\solarmasses{\gtrsim10^5}), formed  within $20$~pc, could spiral  down to the
central  parsec within  the  lifetime of  the  most massive  stars.  It  would
however be  stripped during  its cruise, so  that it would  eventually deposit
only a small fraction of its mass  into the central region. However, it is not
yet clear whether this scenario  could explain the stars closest to \SgrAstar,
at only  a few light days,  and \citet{kim02} and  \citet{kim04} conclude that
they are unable to find realistic initial conditions to their simulations that
would lead to the observed situation through infall of a cluster.

These two possibilities should lead to somewhat different clusters in the end:
for  instance, the  spiralling-in cluster  scenario probably  predicts  that a
significant number of young stars of all masses, co-eval with the inner parsec
early-type stars,  should be found  in the few  central tens of parsecs,  as a
byproduct of the stripping of the  infalling cluster. Such a population is not
yet known.  Actually  observing it would require at least  deep imaging in the
J, H and K  bands, at the best available resolution and  over a field of about
$3\times3$~arcmin$^2$.

The observed properties of the early-type  stars that need to be addressed are
the following:
a group of 15 N- and He-rich stars (the LBVs and the ON stars) are seen within
a small region (about $5\arcsec$ or one light-year), offset from \SgrAstar\ by
about  $1\arcsec$;  there  are  about  two dozen  Wolf-Rayet  stars  dispersed
apparently randomly (in projection) in the central few parsecs; and there is a
compact cluster  of evolved stars  (GCIRS~13E), that contains members  of both
the above-mentioned groups of stars.
These  properties  can  be  classified   in  two  types:  first,  the  spatial
distribution of the stars, and second their spectral types. The actual spatial
distribution  is clearly  subject  to caution  as  we only  have  access to  a
projected  distribution.   However,  at  least  two  velocity  components  are
available  for all  the stars  within the  central two  parsecs, and  a radial
velocity is also  being obtained for a growing number of  them: these four to
five  dimensions should  be  enough  to constrain  the  dynamical models.   An
important point  in the spatial distribution  is that several  groups of stars
are  observed. One  of the  groups, GCIRS~13E,  is a  compact cluster.   It is
proposed in  \citet{maillard04} that  this could be  the remaining core  of an
infalling cluster.  However,  it must be asked whether  such a compact cluster
could itself form \emph{in situ}.

Concerning the spectral  types, the fact reported here  that more stars within
the  GCIRS~16 cluster  are N-  and  He-rich should  also be  addressed by  the
models. These kinds of stars have experienced an unusual mixing. It can be the
trace of a fast initial rotation of  the proto-stars, that would be due to the
initial conditions.  It  may be explained by the shear if  the stars have been
formed within  an accretion  disk.  The efficient  mixing within  the GCIRS~16
stars could also come from  close-by interactions with \SgrAstar, but it would
be hard to explain how these stars can be seen know as an offset group, rather
than a spheroidal cluster centred on \SgrAstar.

\section{Conclusion}

Spectro-imaging techniques have been applied  to the GC, providing very valuable
information on  the nature of the early-type stars lying there,  and on their
spatial  distribution. Upcoming  commissioning of  the adaptive  optics system
that will be used in conjunction  with SPIFFI must be anticipated to give even
more spectacular results.  These stars  appear to belong to several segregated
groups,  distinct in  their spatial  distribution and  stellar types,  but not
apparently in  their age. One  of these groups  is made of  moderately evolved
stars that  show core-processed  material at their  photosphere, which  is not
exceptional, but  unusual. The  more evolved stars,  already at the  WR stage,
have probably lost  any spectral signature that would  have shown whether they
were  already  N- and  He-rich  before they  reached  this  stage.  All  these
observations are  far from trivial and  should give strong  constraints on the
formation scenarios.

\section*{Acknowledgements}

T. Paumard wants to thank  Sergei Nayakshin for fruitful discussions about the
possibility of stellar formation in a parsec-scale accretion disk.

\bibliographystyle{aa} \bibliography{biblio.bib}

  





\end{document}